\documentclass[11pt,twoside]{article}

\usepackage{asp2021}

\aspSuppressVolSlug
\resetcounters

\begin{document}

\title{Combining Data and Metadata: Hybrid Tabular Data Formats}

\author{Mark~Taylor}
\affil{H.~H.~Wills Physics Laboratory, Tyndall Avenue,
       University of Bristol, UK;
       \email{m.b.taylor@bristol.ac.uk}}

\paperauthor{Mark~Taylor}{m.b.taylor@bristol.ac.uk}{0000-0002-4209-1479}{University of Bristol}{School of Physics}{Bristol}{Bristol}{BS8 1TL}{U.K.}


\begin{abstract}
When working with astronomical data, metadata is also important.
A general-purpose file format for transmission, processing and archiving
large datasets should facilitate, among other things,
both efficient processing of bulk data
and encoding of rich semantic metadata.
When choosing a format for a particular purpose sometimes
no existing format satisfies both these requirements adequately,
but combining one data-efficient and one metadata-rich format can be
made to do so.

This paper discusses considerations for designing such hybrid data/metadata
formats, and reviews some examples such as VOParquet, FITS-plus and ECSV.
We focus on tabular data, but some of the considerations may apply
to other datatypes such as arrays as well.
\end{abstract}


\section{Introduction}

Many considerations go into selecting a suitable file format for
a given purpose when working with astronomy data.
These include
availability of compatible software,
ease of use by machines,
ease of use by humans,
efficiency of processing in one or several modes,
storage size,
flexibility,
support for required data types 
and ability to represent rich metadata. 

Ideally, one existing file format presents itself
as satisfying all the required criteria.
Where that is not the case however,
compromises may be made on one or several requirements,
or the decision may be taken to invent a new format
or to adapt an existing one.

One approach in this situation is to take a format that provides
some of the required functionality and combine it somehow with
another format that supplies the missing pieces.
There are several examples in current usage where such a ``hybrid''
format has been defined for use with tabular data in astronomy,
in particular where an efficient but metadata-poor format for
storage of bulk tabular data
(lacking for instance column units and descriptions)
is combined with a
metadata-rich but less performant format for storage of
the associated per-column and per-table semantic metadata,
along with some arrangement to associate the data and metadata parts together.

We discuss here the development and use of such hybrid data/metadata formats.
Section \ref{sec:196-design} gives an overview of design considerations,
section \ref{sec:196-association} explains the options for
associating the data and metadata parts, and
section \ref{sec:196-examples} lists some examples of this approach
in practice.

\section{Design Considerations}
\label{sec:196-design}

When designing a hybrid data/metadata format,
a number of considerations must be taken into account.

Software support is clearly important;
it may be possible to work to some extent with the resulting format using
standard software intended for one or other of the combined formats,
or it may be necessary to write new I/O code from scratch.
For writing the hybrid format, and for reading the data and metadata together,
at least some modifications to standard software are likely to be required.
If the combined format remains comprehensible to off-the-shelf software,
for instance if standard tools that understand the data format
can be used even though they don't take account of the additional
metadata in the hybrid format, utility of the hybrid format is increased.

The data format must be capable of representing the required data,
and the metadata format must be capable of representing the required metadata.
Additionally a good match, or at least compatibility,
between these formats is desirable.
For instance if two formats that both annotate datatypes are combined,
problems can arise if one of the formats supports some datatypes
that the other does not.

It may also be desirable to make the data or metadata readily
readable or editable by humans.

\section{Data/Metadata Association}
\label{sec:196-association}

The chief technical issue in designing a hybrid format is to
define the means by which the metadata and data are joined
into a single unit that can be handled by I/O software.
This is usually done by adapting one of the formats to
``wrap'' an instance of the other.
This can be done either way round.

In some cases, instances of the hybrid format are instances of the data format,
but with some additional metadata (encoded in the metadata format)
stored somewhere in the data format that does not disrupt or invalidate it.
For example a VOParquet file is a Parquet file with a VOTable
header stored in the key-value storage area of its footer.

In other cases, instances of the hybrid format are instances of the
metadata-rich format, but with some arrangement to store bulk data in
the data-efficient format instead of the usual bulk storage mode.
For example a FITS-serialized VOTable is a VOTable file,
but the element which would normally encode the bulk data in
XML elements or a native VOTable binary stream
instead encodes it as a FITS file or binary stream.

If done carefully, this wrapping can be done without disruption to
the ``wrapper'' format, so that such files can be manipulated
by standard software unaware of the hybrid nature of such file instances.

This wrapping can be done by embedding or reference.
FITS-serialized VOTable for instance allows both options:
the VOTable STREAM element that carries the FITS data may either
include the FITS binary stream as a base64-encoded child text node,
or it may reference an external resource containing the FITS payload.
While referencing an external file facilitates manipulation of the
the data and metadata separately, it incurs the responsibility
to keep the two files or resources together which can be problematic;
in most cases the embedding option is more convenient and robust.

\section{Examples}
\label{sec:196-examples}

We review here some examples of hybrid table data formats in current use
in astronomy.
Section \ref{sec:196-component} lists some of the data- and
metadata-oriented formats in use, and
section \ref{sec:196-hybrids} discusses how these have been put together
to form working hybrid data formats.

\subsection{Example Component Formats}
\label{sec:196-component}

The following are some formats suited to storage of bulk data:
\begin{description}
\item[Parquet]
      Apache Parquet is a modern industry-standard format
      optimised to allow fast and
      parallel processing of large and very large tables.
      Many off-the-shelf big data tools can work with it.
      It has extremely limited standard semantic metadata,
      but a key-value list allows insertion of custom metadata items.
\item[FITS]
      The Flexible Image Transport System \citep{2010A&A...524A..42P}
      is a venerable binary format for storage of astronomy array and
      tabular data.
      It is ubiquitous in astronomy, lean, efficient, and easy to implement,
      but clunky, old-fashioned, not suited to parallel processing,
      and has restrictive arrangements for metadata.
\item[CSV]
      Comma-Separated Values is a text-based table format.
      It can be read and written almost anywhere, but is inefficient and
      has no facilities for metadata storage beyond column name.
\end{description}
The following are capable of storing rich metadata:
\begin{description}
\item[VOTable]
      VOTable \citep{2025ivoa.spec.0116O} is an IVOA standard
      XML-based format for tables in astronomy.
      It can annotate tables and columns with units, UCDs,
      coordinate system information, DataLink service descriptors,
      and other items found to be valuable in the context
      of the Virtual Observatory.
\item[YAML]
      YAML Ain't Markup Language is a general purpose text-based format
      capable of storing hierarchical information in a
      human- and machine-readable form.
\end{description}

\subsection{Example Hybrid Formats}
\label{sec:196-hybrids}

The data storage capabilities of FITS have been combined with the
metadata encoding capabilities of VOTable in more than one way.
FITS and VOTable datatypes are compatible by design since VOTable was
intended specifically to represent the same data as FITS binary tables
but with enhanced metadata capabilities,
as well as lifting some other restrictions.
From its inception VOTable has had the option of storing its data
in an embedded or referenced FITS stream with description supplied
by VOTable metadata.
However, this ``FITS-serialized VOTable'' format is rarely used in practice;
if the FITS file is associated by reference the two files are liable
to get separated, but the alternative of embedding the FITS stream
using base64 in the VOTable XML document makes it difficult to
take advantage of the efficiency of the FITS binary format.
An alternative known as ``FITS-plus''
which reverses the direction of wrapping
stores a data-less VOTable document encoding metadata only
in the primary HDU of a FITS file,
so that the FITS file can be manipulated using normal FITS readers,
but FITS-plus-aware readers can benefit from the VOTable metadata
if it is present.
FITS-plus is not currently a public standard, but has long been in use
as a private convention by the STIL library \citep{2005ASPC..347...29T},
and hence the TOPCAT and STILTS applications.

VOTable metadata has also been used in recent years
to enhance the Parquet data format,
since Parquet is gaining considerable popularity for storage
of large astronomy tables but lacks even basic semantic metadata
such as column units.
Like VOTable/FITS, the VOTable/Parquet hybrid also exists in two forms.
The VOTable I/O library in Astropy \citep{2022ApJ...935..167A}
supports a format called \texttt{votable.parquet}
which references an external Parquet file from a metadata-only VOTable
document, in the same way as FITS-serialized VOTable;
however this is outside of the VOTable standard so cannot be read
by standards-compliant VOTable readers, and it also suffers the
inconvenience of splitting the information between two files.
A more recent initiative is VOParquet \citep{196-voparquet}
which like FITS-plus stores a metadata-only VOTable in a permitted
location in a normal Parquet file (the key-value store in its footer).
VOParquet is being adopted by a number of projects
and forms part of the evolving HATS framework \citep{196-hats}
for efficient storage of all-sky datasets using structured Parquet filesets.
The most commonly used Parquet datatypes map straightforwardly on to VOTable
equivalents, but Parquet files can exist which cannot be accurately modeled
with VOTable metadata, so there are restrictions in the applicability
of this pairing.

YAML has also been used to annotate metadata-poor formats.
ECSV (Enhanced Character Separated Values) prepends a YAML header
to a CSV file to provide additional metadata;
the resulting file is neither YAML nor CSV so requires custom
software to read, but this can be fairly easily assembled given
existing YAML and CSV parsers.
MAML is a YAML-based table metadata format that can be stored
in the footer of a Parquet file similarly to how VOParquet uses VOTable.
Since YAML doesn't describe a table format on its own there is no
problem with datatype mismatches,
but this means some custom convention has to be established for
YAML encoding of the per-table and per-column metadata.

\section{Conclusion}

The combination of one data-efficient and one metadata-rich format
has been made several times in astronomy to define a hybrid
storage format for tabular data.
While not the most elegant approach, this can provide a convenient and
effective way to get the best of both worlds
without having to design a new data format from scratch.

\bibliography{196}  


\end{document}